# Single-shot imaging of ultrafast all-optical magnetization dynamics with a spatio-temporal resolution

T. Zalewski and A. Stupakiewicz

*Faculty of Physics, University of Bialystok, 15-245 Bialystok, Poland*

**Abstract.** We present a laboratory system for single-shot magneto-optical (MO) imaging of ultrafast magnetization dynamics with MO Faraday's rotation sensitivity of 4 mdeg/μm. We create a stack of MO images repeatedly employing a single pair of a pump and defocused probe pulses to induce and visualize MO changes in the sample. Both laser beams are independently wavelength-tunable allowing for a flexible, resonant adjustable two-color pump and probe scheme. To increase the MO contrast the probe beam is spatially filtered through a 50 μm aperture. We performed the all-optical switching experiment in Co-doped yttrium iron garnet films (YIG:Co) to demonstrate the capability of the presented method. We determine the spatial-temporal distribution of the effective field of photo-induced anisotropy driving the all-optical switching of the magnetization in YIG:Co film without an external magnetic field. Moreover, using this imaging method, we tracked the process of the laser-induced magnetization precession.

## I. INTRODUCTION

The development of femtosecond lasers and time-resolved methods provided the unprecedented capability for the all-optical coherent control of spins on ultrashort time scales[1]. From an application perspective, the main challenge for obtaining an electronic-competitive solution is the integration of photonics and spintronic devices with triggering of magnetic order by an ultrafast all-optical laser stimulus[2]. The modern magneto-optical methods using pump-probe geometry are attractive tools for research in ultrafast magnetism, offering insight into the time-resolved dynamics of the light-induced processes on ultrashort timescales. The exceptionally sharp time resolution down to femtoseconds offers access to information inaccessible in the slow, static, or quasi-static regime experiments. At the same time, the spatial resolution of the magneto-optical microscopy[3] allows imaging of features as small as couple hundred nanometers, up to the light diffraction limit.

The visualization of the ultrafast dynamics of photo-induced processes in physics is critical for understanding the main interactions. The main advantage of such a technique is compounding both the spatial, and temporal behavior of the phenomena at the same time, giving three-dimensional sets of information. Image, as a 2D representation, gives briefly qualitative information, which by appropriate scaling of intensity level can be quantitive. Imaging methods, which are based on pump and probe techniques are well-known for almost 40 years where the first processes were recorded on photographic films.[4] Later, a huge amount of different variations of this general idea were used very widely, mainly in chemistry and biology.[5,6] However, the pump-and probe methods are inapplicable for all phenomena. Nonrepetable or difficult to reproduce ultrafast processes, such as irreversible chemical reactions, laser-induced damage, or scattering on a living tissue require so-called single-shot imaging. In this context, the "single-shot" term refers to capturing the whole event without repeating it. The whole measurement sequence is performed in real-time with only one laser pulse.[7] In contradiction, it has to be noticed that in this article the discussed method is a repeatable pump and probe technique and the term single-shot refers to the magnetization switching by a single laser pulse.

Recently, different fundamental mechanisms which are responsible for the ultrafast laser-induced magnetization dynamics have been revealed. In the case of the heat load in magnetic materials, they can be separated into thermal and non-thermal. The thermal effect has been observed in metals base on the ultrafast demagnetization during heating close to Curie point.[8–10] The non-thermal effects are caused by the inverse Faraday effect[11] and the photo-magnetic effect.[12,13] Since the discovery of these effects, rapid development, and increased interest in the possibility of magnetization switching by laser pulses in a large class of materials occurred,[14] and time-resolved imaging was a prominent tool for examining the spatio-temporal behavior of the magnetization. Two main approaches for time-resolved magneto-optical imaging can be noted. It can be realized via scanning a focused beam through the sample[15–18] or via direct imaging on a CCD camera[19,20] which is also realized in this work. The method is widely known and it was used for examining plenty of different materials such as GdFeCo metallic alloys,[18,20–22] the rare-earth orthoferrites DyFeO$_3$[23] and HoFeO$_3$,[24] Co/Pt ferromagnetic multilayers,[25] garnets.[16,26] Moreover, the understanding of ultrafast magnetization dynamics was achieved through the development of time-resolved imaging methods using femtosecond sources of X-ray radiation[27] and free-electron lasers.[28] Recently, it was discovered that without any external magnetic field reversible photo-magnetic switching in Co-doped yttrium iron garnet films (YIG:Co) can be obtained only by a single laser pulse.[29] Such switching mechanism is based on the light-induced effective field of photo-induced magnetic anisotropy, which lifetime is about 20 ps, and it is sufficient to trigger angle precession of the magnetization. The photo-magnetic effect is the most intriguing due to the non-dissipative mechanism of all-optical switching of magnetization with the lowest heat load and fastest time switching. No long-timescale recovery from excessive heat load gives a high repetition rate limit of 20 GHz for a cold photomagnetic recording.[30]

Ultimately, imaging with only one ultrashort probe pulse from a laser source is marked with a large number of technical difficulties. These are among others: nonuniformity of the illuminating light, temporal stability of the laser pulse train, separation between pump and probe signal, impact of the diffraction, and mechanical stability. Therefore, to sufficiently investigate the dynamic of such a fast process one has to develop a reliable and repeatable imaging system. For YIG:Co films such measurement has been also presented in[29]. However, obtained image contrast in performed measurements was marked as insufficient, concealing the desired information in a spatially averaged, blurry picture. Detailed spatial analysis and determination of the spatial distribution of the effective field of the photo-induced anisotropy were not performed.

Here, we developed a spatially improved time-resolved system for single-shot imaging of ultrafast laser-induced spin precession and switching. We used a two-color independently tuned laser pump and probe pulses with duration <40 fs at a wide spectral range of 290–2570 nm. Moreover, the two separate mechanical delay lines for both pump and probe beams allowed to ensure full flexibility and independence in settings desired time shift resolution <8 fs. We applied spatial filtering to the unfocused probe laser beam which allows to cleanse the beam of imperfections generated by defects on the sample and the effect of the interference. We demonstrated time-resolved laser-induced dynamics of magnetization precession and permanent switching with a high sensitivity of magneto-optical rotation in YIG:Co thin films. Moreover, we determined the spatial distribution of the field of photo-induced anisotropy driving the magnetization switching.

This paper is organized as follows. In Sec. II, we describe the experimental details of the set-up for single-shot time-resolved imaging, the magnetic structure of the sample, and the image processing. In Sec. III, we demonstrate the imaging of all-optical photo-magnetic switching and precession in YIG:Co film with temporal and spatial resolutions. The conclusions are in Sec. IV.

## II. EXPERIMENTAL DETAILS

The set-up used for the magnetization dynamics imaging is presented in FIG. 1. We employed the time-resolved technique of magneto-optical polarized microscopy using excitation and detection with a single femtosecond pump and probe pulses, respectively. A femtosecond laser system is one-box integrated Ti:Sapphire amplifier (Coherent Astrella). Its output pulse energy is 8 mJ with the 1 kHz repetition rate and pulse duration not longer than 35 fs on 800 nm central wavelength. Next, the laser beam is divided into two branches the probe and the pump beam. The magneto-optical imaging in ferrimagnetic dielectrics requires a two-color scheme between pump and probe due to the spectrally-selective excitation of magnetic order and optimization of magneto-optical rotation for a probing. Two optical parametric amplifiers (Light conversion TOPAS-Prime) and optical frequency mixers (Light conversion NirUVis) were applied to each branch with input pulse energy of 3.5 mJ. The wavelength of both pump and probe pulses can be independently tuned in the 290–2570 nm range.

Afterward, laser pulses are temporarily shifted by Δt defined by the mechanical time-delay line up to 4.4 ns. The probe beam passes through a pinhole PH with a high damage threshold and 50 μm diameter. Next, the beam goes through the polarizer P as a half-wave plate and the adjustable lens with a focus distance of 10 mm. It stays on sample S slightly out of focus to illuminate a sufficiently large area.

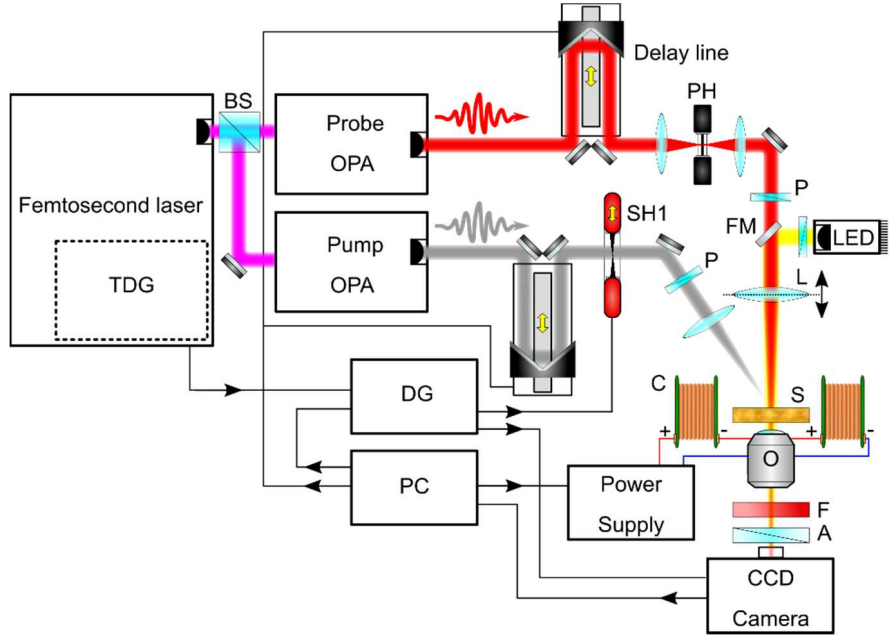

FIG. 1. Magneto-optical setup scheme for the two colors time-resolved single-shot imaging. BS – beam splitter, OPAs – optical parametric amplifiers, PH – pinhole, SH1- mechanical shutter, FM – flip mirror, L – adjustable lens, S – sample, C – coil, F- color filter, A – analyzer, P – polarizers, DG – delay generator.

To retrieve magnetic contrast from images the principle of the magneto-optical polarizing microscope has been used. The light propagating through the sample is affected by the magneto-optical Faraday effect. It induces rotation of the plane of polarization proportional to the magnetization component projected on the direction of the light propagation.[3] In used Faraday (transmission) polarizing microscope geometry, linearly polarized probe pulse propagates through the sample and it is collected by the objective O. Afterward, the pulse passes through the analyzer A. The mutual rotation between polariser and analyzer defines magneto-optical contrast. The CCD camera (Princeton Instruments ProEm 1024) working in the full-frame mode is used to record and digitize 16-bit images with resolution 1024×1024 pixels. To suppressed thermal noise and unwanted charge accumulation, the built-in continuous cleaning of the array procedure was used between each frame. It removes any charge from the array until a trigger pulse is detected and stops as soon as the frame collection begins. The spot size of the probe laser beam localized in the sample surface was 350 μm in diameter which is significantly greater than the pump beam spot size of 130 μm in diameter. For adjusting the probe spot size focusing lens L ($f$ = 100 mm) was used. It makes the beam divergent after its focal point resizing the illuminated area. To eliminate residual pump light the spectral bandwidth filter F was placed before the CCD camera. The pump pulse train is suppressed by the synchronized mechanical shutter SH1. It can release a single pulse to excite the sample. For safety reasons additional mechanical shutter, not marked in the figure was added to the probe branch. The recorded magnetic domain can be removed by an external in-plane magnetic field $H$ produced by coil C.

Moreover, for static imaging and magnetic domain structure observation, the sample also was illuminated by the white LED light let in by a flip mirror FM in normal incidence. To obtain high-quality images of magnetic domains the magneto-optical contrast was optimized by setting the analyzer to respect of polarizer axis in the position which allows visualizing magnetic domains structure. It was done on a long time scale with an LED light source polarized in the same plane as the probe pulse. Next, the intensity of the illuminating probe laser pulses, its uniformity, and angle of incidence were balanced to minimize the total impact of the imaging aberrations, diffraction on surface defects, and created interface fringes concealing the image.

A. Setup synchronization

To automatize and ensure repeatability and stability of the experiment, the following measurement procedure which consists of four integral steps was applied. Firstly, the software sets the mechanical delay line (Newport DL325) to the desired distance, which defines a time delay between the probe and pump pulses. The minimum incremental motion with guaranteed bidirectional repeatability of ±0.15 μm corresponds to about 8 fs resolution, which is shorter than a laser pulse duration. Secondly, the software automatically sends a predefined 100 ms duration voltage pulse to the erasing coil using a power supply (Kepco BOP 72-6 DL). It creates the magnetic field with an amplitude of $H$=100 Oe used to restore the initial state of the sample's magnetization. Next, while the pump beam is still suppressed by the mechanical shutter, only the probe beam illuminates the sample. The image recorded in this step acts as the background image. Ensuring a new background image for every measurement sequence helps to prevent the impact of long-time laser beam instabilities. Finally, the mechanical shutter opens, and both beams – pump and probe illuminate the sample during the single shot of 35 fs creating the time-resolved image. The whole sequence is consistently repeated for every delay value $\Delta t$.

The electrical signal which acts as a trigger is defined by the laser's system regenerative amplifier Astrella. Its rising edge is synchronized with the optical pulse generation (see Fig. 2a). To perform a single-shot imaging experiment in the above sequence, one has to be able to pick only one laser pulse from the pulse train on demand. Plenty of different solutions, mainly dependent on the laser repetition rate, can be applied to solve this issue. The conceptually most simplistic approach is to use the mechanical shutter. Such devices have a large aperture size which, does not affect the beam parameters. However, due to the large inertia of these devices they can be applied only for relatively small (tens of Hz) repetition rates. For higher repetition rates, electro-optic or acousto-optic modulators have to be used. Shutter based on this type of device ensures perfect timing, but they introduce significant losses in light transmission. Moreover in our case, the pulse selection from both pump and probe pulse trains has to be independent.

Significantly longer open-close time cycle (<35 ms) of used mechanical shutter (Thorlabs SHB05T) and several millisecond CCD sensor data acquisition speed makes it necessary to decrease the laser repetition rate. Here, as presented in Fig. 2a, the laser repetition rate was reduced by the amplifier's built-in Pockels cell divider by 40 times, from 1 kHz to 25 Hz. Moreover, to provide a proper pulses synchronization the digital time delay generator DG (Stanford Research Systems DG645) was applied. It is capable to pick not more than one trigger pulse per second, defining the base system operation frequency at 1 Hz (Fig. 2b). This gives a sufficiently long time window for other components and ensures a single pulse selectivity. Because the opening and closing cycle of the mechanical shutter takes about 10 ms the first pulse is always suppressed. Thus, DG always chooses second, subsequent pulse (Fig. 2c). The introduced time shift from the triggering pulse to the sequent optical pulse ensures enough time for opening the mechanical shutter (Fig. 2d). The CCD camera is triggered right after the mechanical shutter. Its exposure time is set to 1 ms, but it is not essential in this type of experiment. The camera requires additional dozen milliseconds for data acquisition.

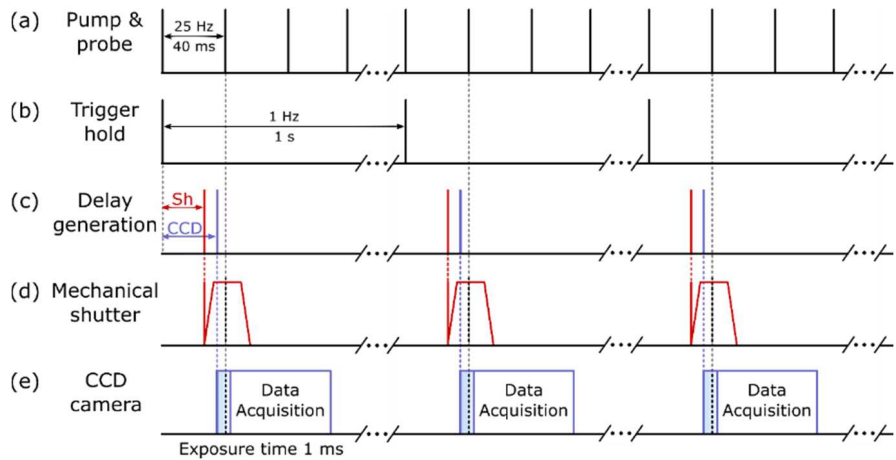

FIG. 2. The timing synchronization scheme. a) The pump and probe pulses are formed from an electronically triggered amplifier's pulse. This signal acts as the synchronizing edge. b) DG holds the trigger for 1 s c) DG omits the first pulse and synchronizes the shutter and CCD camera with subsequent one d) The mechanical shutter open-close time e) The CCD camera exposure and time required for data required. The delay between pump and probe is unnoticeable in the presented scale.

We note that our femtosecond laser system is also capable of working in a single-shot or burst regime. The built-in Pockels cell can be steered directly, deterministically creating the required pulse train. However, every such event disturbs the amplifier's cavity thermal equilibrium. We observed that in long term, it negatively affects the generated pulse train stability. Using external modulators makes it possible to stabilize the laser cavity. Even for the decreased operation frequency, the pulse train energy fluctuates less than in burst or single-shot regimes. The whole system is unified and controlled by the PC using the LabView software. It ensures flexibility and gives the possibility of integrating the system with various other components such as coils, step motors, modulators, or heaters, hence designing many variants of experiments.

## B. Magnetic states in a garnet film

The investigated sample is a 7.5 μm-thick film of Co-doped yttrium iron garnet[31]. The sample was grown by a liquid phase epitaxy method on a gadolinium gallium garnet substrate with 4° miscut. The material is optically transparent in NIR[32] with a relatively low static Faraday rotation of about 0.4° at a wavelength of 650 nm[33]. The garnet has a cubic lattice with two antiferromagnetically coupled spin sublattices of $Fe^{3+}$ in both tetrahedral and octahedral sites. In both of the sublattices, cobalt dopants replace the $Fe^{3+}$ with $Co^{2+}$ and $Co^{3+}$ ions.[34] The addition of cobalt ions introduces strong magnetocrystalline anisotropy and large Gilbert damping α = 0.2.[35] The negative cubic magnetic anisotropy gives 4 easy magnetization axes to be close to the cube diagonals of [111]-type directions (see Fig. 3a) The miscut was implemented to partially break the degeneracy between the magnetization states at room temperature and with no applied field and makes it easier to magneto-optically distinguish the domain structure[29]. The orientations of easy magnetization axes are close to the cube diagonals [111]–type directions. Here, we focus on the simplest case switching between large domains M- and M+, state [11-1] and [1-11] respectively.

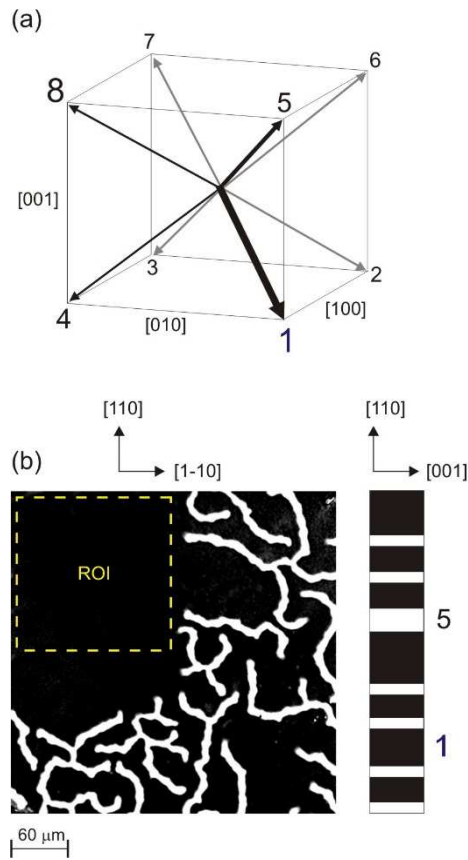

FIG. 3. (a) Easy magnetization axes in YIG:Co film (b) and image of remanent magnetic domain structure in the YIG:Co obtained at zero magnetic fields. The sample was illuminated by the LED with improved contrast by image processing. Two magnetic states are visible: large domain (1) M- along [11-1], and small domain (5) along [111] directions. By applying the 100 Oe external field with 1 s duration in the direction along [110] the small domain pattern can be removed. On certain sample localization, only monodomain (single magnetic phase "1") ROI can be selected. The size of the image is 380x380 μm².

## C. Imaging and data processing

Imaging of the sample illuminated by the probe ultrashort laser pulse is not sufficiently efficient to directly visualize the magnetic domains with low magneto-optical contrast in the YIG:Co. To qualitatively improve the images a spatial filter was introduced into the optical setup. Spatial filter pinholes are useful components for maintaining high beam quality in high-energy pulsed laser systems[36]. Such a filter was used to reduce the high-frequency noise in the probe pulses beam profiles (see Fig. 4a). Also, the spatial shift of illumination connected with the delay line movement is minimized. To couple and decouple the beam through the pinhole two lenses with f = 40 mm were used. The imaging beam is not a plane wave, therefore the light profile is not uniform. However, the presence of the magnetic domain structure can be easily distinguished on the images and on the profile of the cross-section (see Fig. 4b).

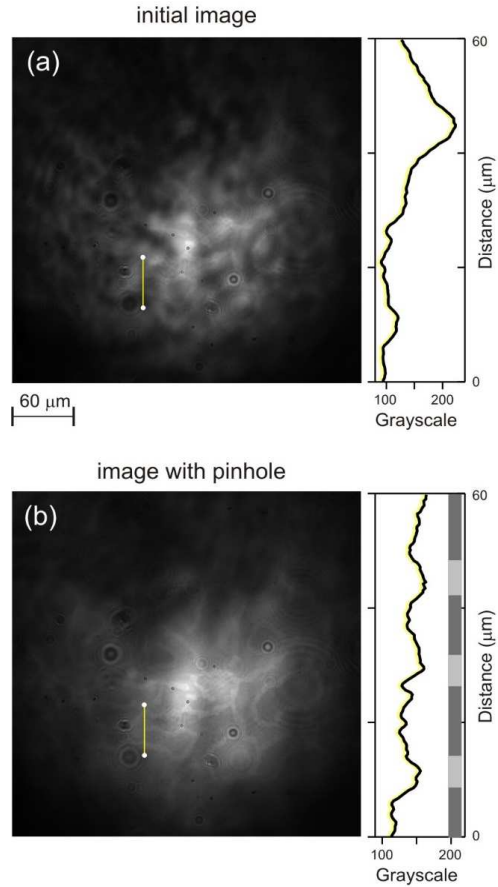

FIG. 4. The magneto-optical images in YIG:Co film which is illuminated with a single probe pulse. The distance profiles correspond to the yellow cross-section line marked in the images. a) initial image was taken without a pinhole. Nonuniformity of the incident wavefront creates high-frequency noise concealing the magnetic structure character. b) image obtained using a pinhole. As in FIG. 3 revealed domain structure corresponds to the large (1) M- and small (5) M+ magnetic phases schematically marked as stripes on the profile. Images are represented in a normalized 8-bit grayscale. The images size is 380×380 μm$^2$.

To determine the character of pump-induced changes from the recorded 2D images, one has to separate the contribution of the magneto-optical Faraday effect from pure optical effects. Any local light intensity changes such as probe beam instabilities or noise caused by the pump absorption in the sample have an impact on the recorded image. Here, the visible difference comes from the rotation of the polarization plane of the probe beam, which passed through the sample. While propagating through the sample, the probe beam interacts with the magnetic domain structure. Different orientation of the magnetization component of magnetic domains results in the different Faraday rotation of the polarization plane, which is detected on the camera. The position of the analyzer defines the magneto-optic contrast.

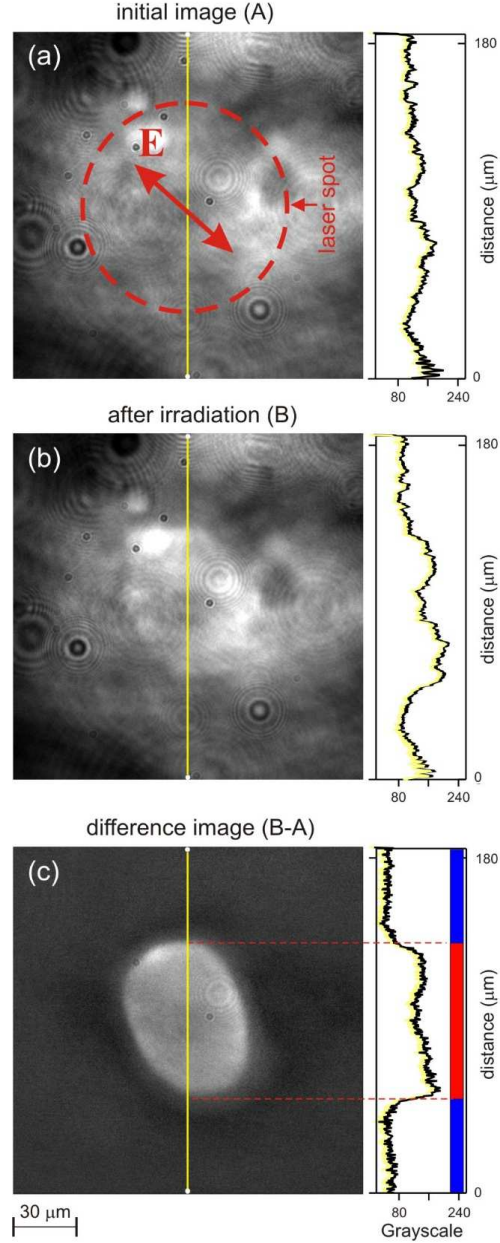

FIG. 5. a) The single-shot imaging of YIG:Co at monodomain state with magnetization (1) M- along the [11-1] direction before excitation, illuminated only by a single probe pulse with λ = 650 nm. b) The image recorded after a single pump pulse with λ = 1300 nm for Δt= 185 ps. The magnetization of the switched domain was along [1-11] direction and corresponds to a state (8). c) The differential image (B-A). The pump polarization was **E**∥ [100]. The slightly elongated shape of the switched domain is related to both the beam shape and the miscut angle of the sample and schematically marked (red dashed line) on a profile.

The images of the domain structure can be visualized directly using only a single probe pulse. It is also valid to the dynamic situation within pump and probe pulses. However, for contrast improvement of these images and to suppress the impact of the probe's wavefront nonuniformity differential images were created. These images were obtained by subtracting image B recorded after single pump pulse irradiation and background initial image A (see Fig. 5).

The most distinct advantage of imaging is that it reveals the spatial information of the domain, concerning its shape, structure, and size. The time-resolved addition to imaging allows determining the time scale of pump-induced spatial changes. Therefore, creating repeatable series of images, as was previously proposed in the measurement sequence, is an easy way to retrieve the information concerning the time and localization of pump-induced changes. Such an image stack contains information about the relative time delay between images, which is retrieved from the delay line distance.

Spatial information can be obtained directly from a stack of images, but several factors need to be considered. Firstly, to compare images obtained with non-constant illumination conditions, one has to uniformize their intensity levels. In order not to spoil the information concerning the single image intensity levels, the contrast-enhancing procedure should be applied to all images in the same manner. Therefore, the histogram of the whole stack, instead of individual images histograms, was used to rescale the intensity levels of each image. Here, float 32-bit representations of differential, images were rescaled to 0-256 range. Secondly, one has to be sure that the background intensity level is flat. For every image, it has to be at the same, uniform level. We note that without it, differential images were spatially incomparable. In this case, we used additional post-processing algorithms.[37]

## II. RESULTS

We used a setup for single-shot time-resolved magneto-optical imaging to determine the photo-magnetic switching in YIG:Co film. The experimental data of photo-magnetic switching retrieved from the differential image stack were represented in a three-dimensional manner, highlighting the mutual temporal and spatial behavior of the change of magnetization orientation as shown in FIG. 6. The magneto-optical differential images for selected Δt are shown on the top panel in FIG. 6. The contrast in these images is due to the magneto-optical Faraday effect, which is proportional to the perpendicular magnetization component. Change of intensity within the laser pump spot on these images allows tracking the evolution of the photo-magnetic switching. To visualize the spatio-temporal redistribution of magnetization the 3D map was created from the intensity profiles of 150 differential images which were recorded using a single probe pulse for different Δt. The intensity profile for every image, proportional to the time-resolved Faraday rotation, was used data for the 3D map. To improve the signal-to-noise ratio, instead of only one profile cross-section line, we chose the averaged profile from a rectangular selection of 10 μm width as shown by yellow on the image in Fig. 6. The color code on the map corresponds to the out-of-plane magnetization component. In this map, we observe the red area of photo-magnetic switching in YIG:Co between M- and M+ states which correspond to magnetization states (1) and (8) in FIG. 3a.

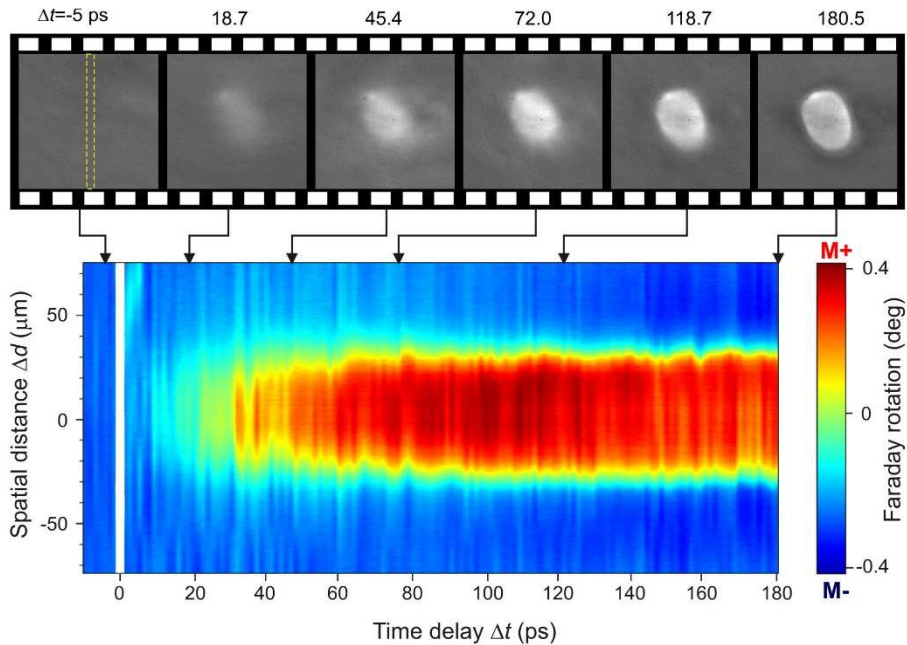

FIG. 6. Three-dimensional representation of magnetization dynamics under single pump pulse with a fluence of 75 mJ/cm$^2$ in YIG:Co film. A negative time corresponds to the probe pulse illuminating the sample before the pump pulse. The characteristic blink in overlap position defines precisely Δt=0 ps time. The top panel shows the differential images of magnetic domain structure obtained at various Δt. The color on the diagram corresponds to the angle of time-resolved Faraday rotation which is proportional to the change of the out-of-plane magnetization component. The deep blue color is an initial magnetization state (1) M-. The green color corresponds to the magnetization aligned in-plane. The red color shows the area switched to state (8) M+.

The mechanism of presented photo-magnetic switching in YIG:Co film is triggered via the precession of the net magnetization.[29] In this case, the pump light optically excites one of the possible and the most effective electronic transitions for Co ions being in the tetrahedral sites.[38] Such excitation is responsible for the change of magnetic anisotropy in a garnet. For an initial state, before pumping the intrinsic magnetic anisotropy is predominantly cubic with a small uniaxial contribution. The strength of the effective field of photo-induced anisotropy is comparable to the intrinsic one.[13] It can lower the energy barrier allowing the magnetization to switch to the second state. Because of the large damping, the lifetime of the photo-induced anisotropy is low and it decays at about 60 ps at room temperature which corresponds to the quarter of the laser-induced precession.

The size of a switched domain can be determined by the spot of the pump beam. Here, it is given by the averaged laser optical power and total pump spot size on the sample, determined by the system focusing. However, because the incident beam has a Gaussian shape, locally the light intensity differs. It is highest on the beam axis, and it decreases away from the axis. Therefore, by selecting and analyzing a selected region of interest (ROI) one can distinguish the amplitude of photo-induced field-related temporal changes in FIG. 7a. Improved sensitivity of the set-up allowed to visualize and distinguish four characteristic regions with different behavior of the magnetization vector. The mean value from the selected ROIs is subjected to further analysis. Chosen ROIs are marked on the image with appropriate colors. All of them are circles with the same size of about 10 μm diameter and are shifted from the image center by given $\Delta d$. Choosing a larger region instead of only individual pixels can reduce noise impact. For the $\Delta d > 60$ μm ROI corresponds to the region which is unaffected by the pump pulse so it acts as a background. No change of magnetization out of plane component appears here.

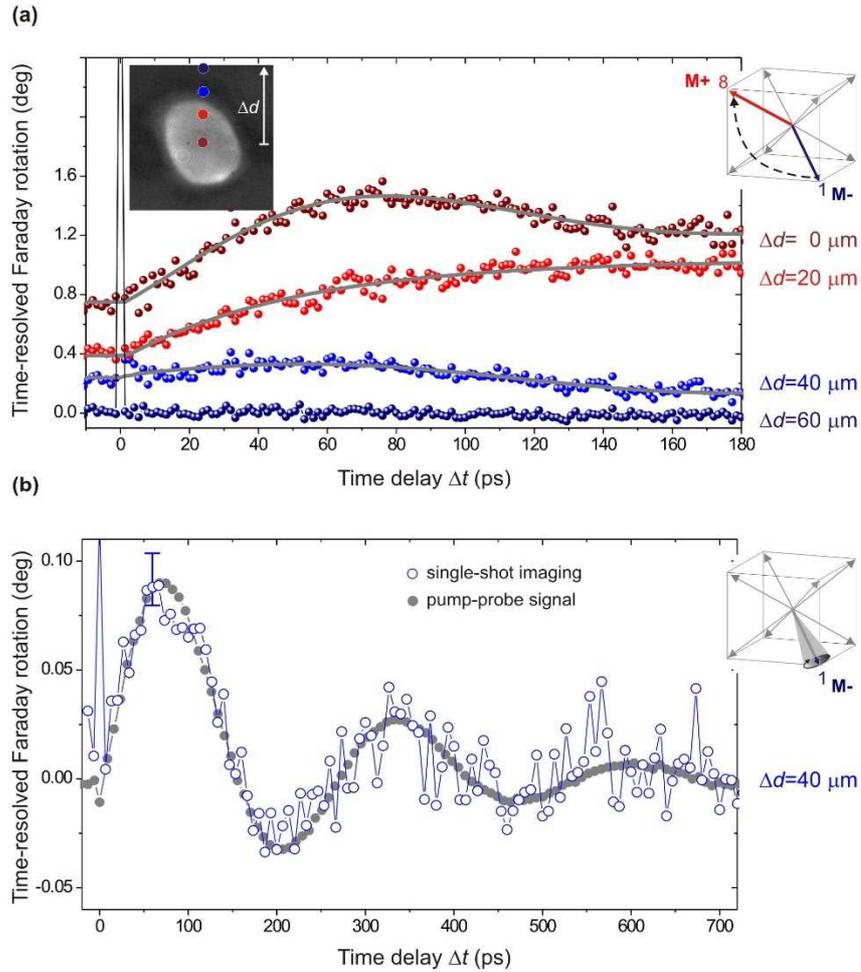

FIG. 7 a) The temporal behavior of the out-of-plane magnetization component for the different characteristic regions of interest. b) The time-resolved precession around the state *M-* in the long-time domain for $\Delta d = 40$ μm. The signal obtained by the single-shot imaging is compared to the regular pump-probe dynamic with the same experimental condition of the laser pulses.

The background is used as the image stack reference. It has to be flat to ensure that all images taken in the measurement series are comparable. For smaller $\Delta d$ = 40 μm the incident pump intensity is still below the switching threshold. However, its intensity is high enough to trigger the precessional movement of magnetization. With $\Delta d$ = 20 μm and lower the incident pump intensity is high enough to obtain the switching of magnetization from state *M-* to state *M+*. After it, the out-of-plane component of magnetization is reversed. Presented behavior applies also to the averaged situation – ROI including whole switched spot. In the center of the spot $\Delta d$ = 0 μm the local pump intensity is highest. It triggers the magnetization switching from one state to another. Previously, in reference[29] due to the insufficient magneto-optical contrast, this effect was hidden. Here, because of the greater imaging sensitivity, we can perform spatial analysis in detail. Thus, for a sufficiently large field of photo-induced anisotropy switching from the state (1) to (8) occurs. Next, after about 60 ps, when the magnetization was switched, we observed damped precession around the direction of the switched magnetization state. The amplitude of this precession corresponds to the FMR mode was small and the magnetization remains at the switched state. This effect was predicted[38] but it was not observed experimentally.

We compared precession for $\Delta d$ = 40 μm measured through presented single-shot imaging with the signal obtained through highly sensitive pump-probe measurement (see Fig 7b). By that, we determined the time-resolved Faraday rotation sensitivity of our method as 30 mdeg for the 7.5 μm garnets which can be scaled to about 4 mdeg/μm. The data presented as the pump-probe signal was obtained within the same setup but with a typical detection pump-probe scheme.[39] The laser system worked with a repetition rate of 1 kHz. The probe beam was modulated by a chopper to 500 Hz and it was focused on the sample to 50 μm in diameter. Next, the probe was directed to the half-wave plate and the Wollaston prism which separates it into two linearly polarized beams with orthogonal polarization. These beams are directed to the two separate branches of an auto-balanced photodiode, which monitors the change of their intensities corresponding to the Faraday rotation. The frequency modulation alternating the pump and probe signal was applied to use lock-in amplifier (Zurich Instruments MFLI) based detection. Moreover, to improve the signal-to-noise ratio boxcar integrator (Stanford Research Sr 250) was applied.

III. CONCLUSIONS

We developed and implemented an automated setup for the high-contrast time-resolved single-shot imaging of magnetization dynamic. By using two OPAs we obtained independent temporal and spectral tunability of two beams. The possibility of selecting the central wavelength from the wide spectrum for the pump beam is a key for inducing different resonant transitions responsible for the magnetization switching, especially in dielectrics. The spectral tuning of the probe beam is necessary to find the wavelength assuring an optimal balance between high magneto-optical contrast and low absorption in the sample. Using two delay lines made it possible to set a pump delay with an unchanging probe delay and vice versa. By applying the pinhole into the probe beam we limited the interference noise improving the imaging quality. We observed the switching in single-domain structure and spatially analyzed its dynamics. Moreover, right next to the switching with the high sensitivity of 4 mdeg/μm, we distinguished regions in which precession of the magnetization appears. The proposed imaging method is characterized by full flexibility and very high sensitivity. It may be further use for examining the spatio-temporal behavior of more sophisticated, multi-state magnetic structures, as well as even finding completely new switching mechanisms in different materials.

**Acknowledgments.** The authors thank D. Afanasiev and A.V. Kimel for the fruitful discussion. This work has been funded by the Foundation for Polish Science POIR.04.04.00-00-413C/17-00.